\documentclass[prl,twocolumn,aps,amsmath,amssymb,showpacs,superscriptaddress]{revtex4}

\usepackage{graphics}
\usepackage{dcolumn}
\usepackage{bm}
\usepackage[T1]{fontenc}

\begin{document}

\title{0.7 Structure and Zero Bias Anomaly in Ballistic Hole Quantum
Wires}

\author{R. Danneau}\email[Corresponding author; r.danneau@boojum.hut.fi]{}
\affiliation{School of Physics, University of New South Wales,
Sydney 2052, Australia}
\affiliation{Low Temperature Laboratory,
Helsinki University of Technology, Espoo, Finland}
\author{O. Klochan}
\author{W. R. Clarke}
\author{L. H. Ho}
\author{A. P. Micolich}
%\author{J. W. Cochrane}
\author{M. Y. Simmons}
\author{A. R. Hamilton}
\affiliation{School of Physics, University of New South Wales,
Sydney 2052, Australia}
\author{M. Pepper}
\author{D. A. Ritchie}
\affiliation{Cavendish Laboratory, J.J.Thomson Avenue, CB3
OHE Cambridge, United Kingdom}

\begin{abstract}

We study %the magnetic field dependence of
the anomalous conductance
plateau around $G = 0.7(2e^{2}/h)$ and the zero-bias anomaly in
ballistic hole quantum wires with respect to in-plane magnetic
fields applied parallel $B_{\parallel}$ and perpendicular
$B_{\perp}$ to the quantum wire. As seen in electron quantum wires,
the magnetic fields shift the 0.7 structure down to $G =
0.5(2e^{2}/h)$ and simultaneously quench the zero bias anomaly.
However, these effects are strongly dependent on the orientation of
the magnetic field, owing to the highly anisotropic effective
Land\'{e} \emph{g}-factor $g^{*}$ in hole quantum wires. Our results
highlight the fundamental role that spin plays in both the 0.7
structure and zero bias anomaly.

\end{abstract}


\maketitle

One-dimensional (1D) systems have revealed some of the most
intriguing effects in solid state physics \cite{giamarchi}. For
example, it is well known that the Coulomb interactions in 1D
electrons systems lead to a correlated non-Fermi liquid state known
as Tomonaga-Luttinger liquid \cite{giamarchi}. Van Wees \emph{et
al.} \cite{vanwees1988} and Wharam \emph{et al.} \cite{wharam1988}
demonstrated the underlying quantum nature of 1D non-interacting
electronic systems, following the seminal work of Thornton \emph{et al.}
\cite{thornton86}. In these experiments it was shown that the
conductance of a ballistic 1D system is quantized in units of $G_{0}
= 2e^{2}/h$ corresponding to the discreteness of the number of
propagating modes in the channel (see \cite{beenakker1991} for a
review). This phenomenon has been studied extensively in 1D electron
systems leading to the discovery of more exotic phenomena such as
the unexpected feature around the unquantized value of $0.7G_{0}$. This conductance anomaly was
pointed out first in the early nineties \cite{patel1991} and studied
later in detail \cite{thomas1996}, the origin of which is still
debated \cite{fitzgerald2002}. Because the non-interacting particle
theory does not expect such a feature \cite{beenakker1991}, this
conductance anomaly, known as the 0.7 structure, is believed to be a
many-body effect. In this work, we present a study of the 0.7 structure and its
non-linear counterpart, the zero-bias anomaly (ZBA),
in hole quantum wires, and demonstrate that both show an anisotropic response to
an external magnetic field, which suggests that both are closely linked and related to spin.

The 0.7 structure is a resonance-like feature that sits below the
first conductance plateau. It appears to be a zero-field remnant
feature of the first spin-resolved plateau $e^{2}/h$ when the spin
degeneracy is lifted by the Zeeman effect \cite{thomas1996}. This
picture is consistent with the fact that the 0.7 structure shifts to
$e^{2}/h$ with the application of an in-plane magnetic field.
Temperature dependence studies showed that while the quantized plateaus are
thermally smeared, the 0.7 structure becomes stronger
\cite{vanwees1991,thomas1996}. This indicates that the 0.7 structure
is not due to electron backscattering from a defect in the vicinity of
the constriction \cite{beenakker1991}, but is thermally activated and
therefore not a ground state property \cite{kristensen2000}. In
addition, the 0.7 structure coincides with an enhanced conductance
at zero source-drain bias, which falls away rapidly as the bias
between source and drain is increased. This conductance peak shares
many of the features of the ZBA in quantum dots
\cite{glodhabergordon1998}, including the splitting and ultimate
destruction of the zero-bias peak with the application of an
in-plane magnetic field. This has led to a description of the 0.7
structure in terms of a Kondo-like correlated state within the
constriction \cite{cronenwett2002,meir2003}.

While debate continues as to the exact origin of the 0.7
structure \cite{thomas1996,cronenwett2002,meir2003,wang1996,seelig2003,sushkov2001,spivak2000},
the anomalous conductance plateau is commonly believed
to be linked with spin. We therefore expect that the magnetic field dependence
of both the conductance plateau and ZBA will strongly depend on
the 1D effective Land\'{e} \emph{g}-factor $g^{*}$, which
determines how the spin couples with an external magnetic field.
Here, we demonstrate this dependence explicitly by examining the
magnetic field dependence of the 0.7 structure in hole quantum
wires. Unlike electron 1D systems which have an isotropic $g^{*}$,
the \emph{g}-factor in hole quantum wires is highly anisotropic
\cite{danneau2006a}. We show that the conductance plateau and ZBA
also share this anisotropy by examining their behavior with respect
to in-plane magnetic fields applied parallel $B_{\parallel}$ and
perpendicular $B_{\perp}$ to the wire. In doing so we highlight the
fundamental role that spin plays in these phenomena.

\begin{figure}[htbp]
\scalebox{0.445}{\includegraphics{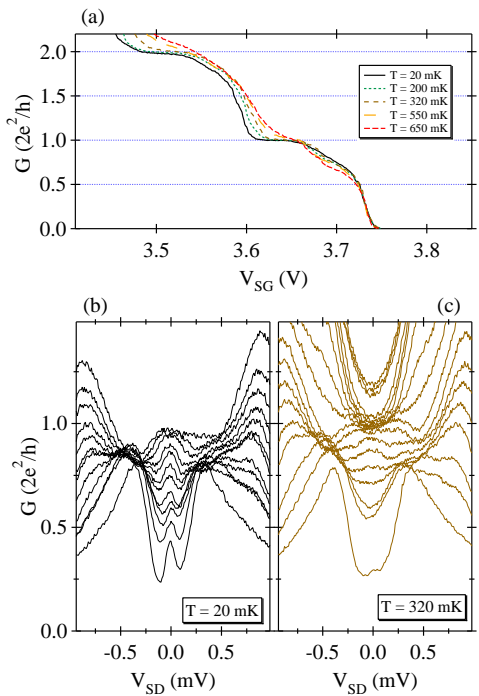}} \caption{(a)
$G$ of the quantum wire versus side gate
voltage $V_{\mathrm{SG}}$, for \emph{T} = 20, 200, 320, 550 and
650 mK; Each curve is shifted to sit on top of the 20 mK curve;
(b) $G$ versus source-drain bias $V_{\mathrm{SD}}$ for different
$V_{\mathrm{SG}}$ at \emph{T} = 20 mK; (c) $G$ versus
$V_{\mathrm{SD}}$ for different $V_{\mathrm{SG}}$ at \emph{T} =
320 mK; data are taken for back and middle gates fixed at 2.5
V and -0.225 V respectively. No magnetic field is applied to the
system; (a), (b) and (c) are data taken from the same cool down.}
\end{figure}

For these experiments, we used the 1D hole bilayer system grown on
(311)A n$^+$-GaAs substrate described previously \cite{danneau2006}.
Electrical measurements were performed in the top wire of the
bilayer hole system (the top layer has mobility 92
m$^{2}$V$^{-1}$s$^{-1}$ and density 1.2$\times$10$^{15}$ m$^{-2}$)
in a dilution refrigerator using standard low-frequency ac lock-in
techniques with an excitation voltage of 20 $\mu$V at 17 Hz. Side
gates and a middle gate created by standard electron beam
lithography define the quantum wires along the $[\overline{2}33]$
direction.

\begin{figure}[htbp]
\scalebox{0.5}{\includegraphics{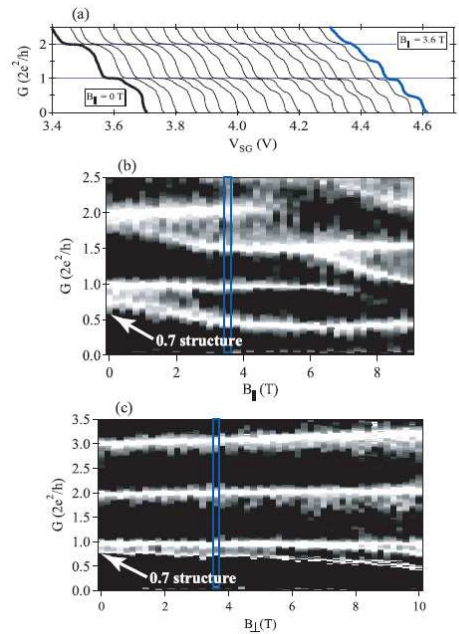}} {\caption{(a)
$G$ of the quantum wire versus side-gate voltage $V_{\mathrm{SG}}$ for
different in-plane magnetic fields parallel to the wire
($B_{\parallel}$) from 0 T (i.e. when 1D subbands are degenerate), to 3.6 T (i.e. when 1D subbands
are completely spin resolved), in
increments of 0.2 T (from left to right). \emph{T} = 20 mK, back and
middle gates are at 2.5 V and -0.225 V, respectively. All curves are offset for clarity.
(b): grayscale of the transconductance as a function of $B_{\parallel}$ up to 8.8 T
and $G$. White regions correspond to low transconductance
(conductance plateaus). (c): grayscale of the transconductance as a function of versus
$B_{\perp}$ and $G$, with similar experimental conditions.
}}
\end{figure}

In Figure 1(a) we show the differential conductance $G = dI/dV$ (corrected for a
2.5 k$\Omega$ series resistance) of the two first 1D subbands
as a function of side-gate voltage $V_{\mathrm{SG}}$ for different
temperatures. The anomalous plateau at 0.7$G_{0}$ becomes
significantly stronger as the quantized plateaus are washed out with
increasing temperature, as has been observed in induced hole
quantum wires \cite{klochan2006}. This indicates that the observed
feature is not a conductance transmission resonance due to a defect
in the vicinity of the constriction. The higher plateaus become
unresolvable above $T \gtrsim$ 600 mK, i.e. at temperatures
approximately an order of magnitude smaller than for 1D electron
systems \cite{thomas1996,kristensen2000}. This is due to the greater
effective mass of holes $m^{*}_{h} \approx$ 5$m^{*}_{e}$, which
results in significantly smaller subband spacings in p-type 1D systems
\cite{danneau2006}. In Figures 1(b) and 1(c) we also show the
conductance as a function of source-drain bias (SDB) for several
$V_{\mathrm{SG}}$. The data show a well defined ZBA (the narrow peak
in conductance around $V_{\mathrm{SD}} = 0$ V) for conductances
below $2e^{2}/h$ [Fig. 1(b)]: For $T \approx$ 300 mK, the ZBA has
vanished [Fig. 1(c)], a temperature which is approximately two times
less than for electron systems \cite{cronenwett2002,note0}. These results
are consistent with previous studies of spin-dependent focussing in
1D hole systems \cite{rokhinson2006}.

\begin{figure*}[htbp]
\scalebox{0.55}{\includegraphics{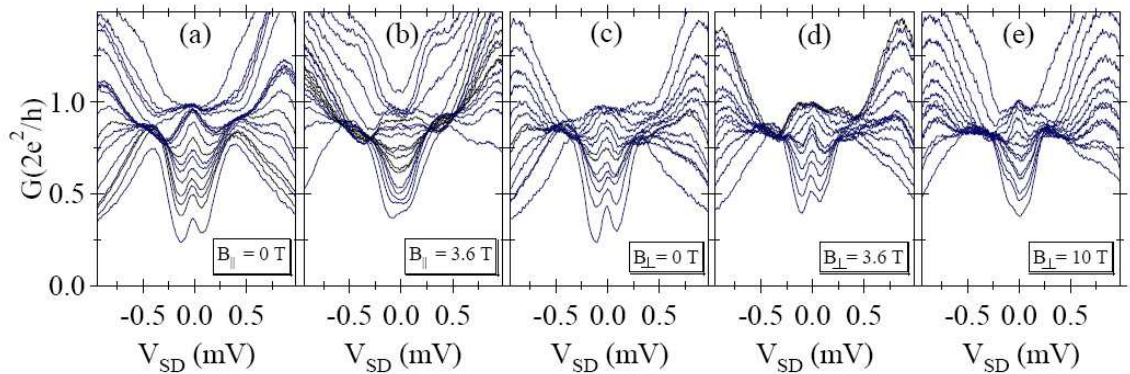}} {\caption{
%Differential conductance $G$
%in the non-linear (finite dc-bias applied) as a function of an in-plane magnetic field $B$
%parallel ((a) and (b)) and perpendicular ((c), (d) and (e)) to the wire.
(a) and (b): $G$ versus
$V_{\mathrm{SD}}$ for different $V_{\mathrm{SG}}$ of the same
quantum wire at (a) $B_{\parallel} = 0 $ T and (b) $B_{\parallel}
= 3.6 $ T, \emph{T} = 20 mK, back and
middle gates are at 2.5 V and -0.225 V, respectively ((a) and (b)
are taken at the same cool down).
(c), (d) and (e): $G$ versus $V_{\mathrm{SD}}$, for different
$V_{\mathrm{SG}}$, for $B_{\perp}$ = 0, 3.6 and 10 T, in similar
experimental conditions ((c), (d) and (e) are taken at the same
cool down; (c) and Fig. 1 (b) are the same data): the ZBA
is clearly observed at $B_{\perp}$ = 0 and 3.6 T. At $B_{\perp}$ =
10 T, the ZBA vanishes at low conductance.}}
\end{figure*}

Once the 1D constriction is created, increasing $V_{\mathrm{SG}}$ moves the Fermi level through the 1D spin
degenerate subbands \cite{beenakker1991}. An in-plane magnetic field
lifts the spin degeneracy of the 1D subbands \cite{wharam1988}, with
the non-degenerate subbands becoming quantized in units of $e^{2}/h$
above a critical field $B_C$. Figure 2(a) shows the effect of the
in-plane magnetic field parallel to the wire $B_{\parallel}$ on the
quantized steps and the 0.7 structure. We observe clearly the Zeeman
splitting of the 1D subbands and the smooth evolution
of the plateau at 0.7$G_{0}$ to 0.5$G_{0}$ = $e^{2}/h$ at
$B_{\parallel}$ = 3.6 T, i.e. to the complete spin-resolved
conductance. Figure 2(b) shows a grayscale of the transconductance
$dG/dV_{\mathrm{SG}}$ of the same data highlighting the shift of the
0.7 structure to 0.5$G_{0}$ around $B_{\parallel}$ = 3.6 T.
After thermal cycling and reorientation of the sample, an in-plane
magnetic field perpendicular to the constriction $B_{\perp}$ was
applied. Figure 2(c), shows a grayscale of the
transconductance at $T$ = 20 mK as a function of $G$ and in-plane
magnetic field $B_{\perp}$ up to 10 T. $B_{\perp}$ does
not affect the quantized steps. This highlights the extreme anisotropy of
the $g$-factor in 1D hole systems that has been observed previously
\cite{danneau2006a}. This anisotropic behavior of the Zeeman
splitting arises from the 1D confinement in this system with strong spin-orbit (SO) coupling
which forces the quantization axis of $\hat{J}$,
originally perpendicular to the 2D plane \cite{winkler2003},
to align along the length of the wire \cite{danneau2006a}.

Consistent with the behavior of the 0.7 structure, SDB measurements show that the ZBA
exhibits the same remarkable anisotropy, demonstrating the
peculiarities of holes compared to electrons.
Figures 3(a) and 3(b) show the conductance as a function of SDB for
$B_{\parallel}$ = 0 T and B$_{\parallel}$ = 3.6 T respectively. The
clear ZBA measured at zero magnetic field is suppressed by
$B_{\parallel}$ = 3.6 T, coinciding with the conductance plateau
reaching $G = 0.5G_{0}$ and fully spin-resolved of higher
subbands. The magnetic field dependence is
significantly different for the $B_{\perp}$ case.
Figures 3(c), (d) and (e) present SDB data from for
$B_{\perp}$ = 0, 3.6 and 10 T respectively, taken during the same
cool down as in Fig. 2(c). The clear and reproducible ZBA remains intact at
$B_{\perp} = 3.6$ T and does not seem to be affected by the
magnetic field. At $B_{\perp}$ = 10 T, the ZBA is finally
destroyed.

Confinement and SO coupling alter drastically the hole properties in zinc-blende
compounds. The 2D confinement lifts the heavy hole (HH)- light hole
(LH) degeneracy with HH's occupying the lowest energy state. This
means that the carrier transport is
predominantly via the HH subband \cite{winkler2003}. It is therefore
surprising that the 0.7 structure evolves smoothly to 0.5$G_{0}$
with increasing $B_{\perp}$ despite no evidence for Zeeman splitting
in the higher subbands. Nonetheless, its evolution is still highly
anisotropic. For the $B_{\parallel}$ case, the 0.7 structure falls
rapidly, reaching 0.5$G_{0}$ at $B_{\parallel}$ = 3.6 T. On the
contrary, the 0.7 structure is barely affected at $B_{\perp}$ = 3.6 T
and reaches 0.5$G_{0}$ only at much larger magnetic fields
$B_{\perp}$ = 10 T. We notice that the evolution of the 0.7 structure
is complicated at very high fields by a weak additional feature
in the conductance that appears to split off from the 1D
subband \cite{note}. However, SDB measurements at
$B_{\perp} = 10$ T reveal that the ZBA is absent at this field,
suggesting that 0.7 structure has indeed reached its spin-resolved
state (i.e. 0.5$G_{0}$). This highly anisotropic behavior with respect
to the in-plane magnetic field reinforces strongly the hypothesis
that \emph{the 0.7 structure and the ZBA are linked and
spin-related phenomena}.
However, the question still remains as to
why we observe any effect of B$_{\perp}$ on the 0.7 structure and
ZBA at all up to 3.6 T, a magnetic field sufficiently large that is leads to a fully spin resolved
state in the parallel orientation?

The higher subbands show little evidence of spin splitting up to
$B_{\perp}$ = 10 T, indicating that the $g^{*}_{\perp}$ for the
first subband is at least 4.5 times less than $g^{*}_{\parallel}$
\cite{danneau2006a} i.e. $g^{*}_{\parallel}/g^{*}_{\perp} > 4.5$. On
the other hand, the shift in the 0.7 structure and the destruction
of the ZBA at $B_{\perp}$ = 10 T suggest that these features are
more sensitive to an external magnetic field. In addition, if we
compare the rate at which the 0.7 structure evolves towards
0.5$G_{0}$ as a function of $B_{\parallel}$ and $B_{\perp}$ we find
that the anisotropy can be \emph{no greater} than
$g^{*}_{\parallel}/g^{*}_{\perp} \leq 4$. This reduction in the
anisotropy is also consistent with calculations based on
our device structure (i.e. quantum well width $W_{x}$ and 2D hole
density) and crystallographic orientation \cite{zuelike2006}.
The calculations predict that 1D confinement induced HH-LH band
mixing results in a significant enhancement of both
$g^{*}_{\parallel}$ and $g^{*}_{\perp}$. However, $g^{*}_{\perp}$ is
predicted to increase more rapidly with confinement, resulting in a
reduction in the $g$-factor anisotropy: our measurements are consistent
with this. With increasing confinement, the
\emph{g}-factor anisotropy is predicted to become zero and then
ultimately reverse in the extreme 1D limit where the width of the
wire $W_{y}$ becomes comparable to the width of the quantum well
$W_{x}$ i.e. $g^{*}_{\perp} > g^{*}_{\parallel}$ for $W_{x}/W_{y}
\gtrsim 0.42$. The continuous transition in \emph{g}-factor
anisotropy corresponds to a monotonic change in the carriers from HH
character in the 2D limit to LH character in the symmetric wire
limit $W_{y}$ = $W_{x}$, which results from confinement induced
HH-LH band mixing.

The extreme 1D limit is inaccessible in our samples due to the
long Fermi wavelength $\lambda_{\mathrm{F}}$. At the first plateau $W_{y} =  \lambda_{\mathrm{F}}^{1\mathrm{D}}/2$ and we
expect $\lambda_{\mathrm{F}}^{1\mathrm{D}}$ to be approximately four times greater than the 2D
Fermi wavelength \cite{williamson1990} i.e.  $\lambda_{\mathrm{F}}^{1\mathrm{D}}/2 \approx$ 140 nm
giving $W_{x}/W_{y} \approx 0.15$ \cite{note1}. However, even at this level of confinement
we can expect to see the effects of band mixing below the first
subband \cite{zuelike2006} and hence a reduction in \emph{g}-factor anisotropy for
the 0.7 structure \cite{note2} - as observed in our experiments.
We note that since the magnetic length $l_{B}=\sqrt{\hbar/eB}$ can be comparable
to the width of the wire even at moderate fields, magnetic confinement may contribute
to the $g$-factor anisotropy at large $B$ (in addition to LH-HH mixing).
However, the anisotropic behavior of the 0.7 structure at low fields is unlikely due to magnetic confinement.
Furthermore since we compare the measured anisotropy of the ZBA and the measured behavior of the 0.7 structure,
the conclusion that both conductance anomalies are spin related is still valid, even with strong SO coupling.

There is not yet a consensus as to the origin of the 0.7 structure. Some of the
theories include spontaneous spin polarization \cite{wang1996},
Kondo effect \cite{meir2003}, electron-phonon scattering \cite{seelig2003}, charge density waves \cite{sushkov2001} or
Wigner crystals \cite{spivak2000}.
While a detailed comparison of our data with these various models is beyond the
scope of this work, our data provide valuable new constraints on theory: while
interactions are strongly enhanced due to the large effective mass, holes
have an effective spin $J = 3/2$, so the spin-physics could be different.

To summarize, we have studied the 0.7 structure and the ZBA in 1D
hole systems. We have shown that both features react to an in-plane
magnetic field depending on its orientation with respect to the
quantum wire direction. These findings strongly support, firstly
that \emph{0.7 structure and ZBA are directly linked}, and secondly
that they are \emph{both related to spin}. Finally, we have shown
that the 0.7 structure evolves slowly to 0.5$G_{0}$ despite the
suppression of the Zeeman splitting for HH when an in-plane magnetic
field is applied perpendicular to the wire: this may be the result of HH-LH
mixing induced by the 1D confinement.

We wish to acknowledge U. Z\"{u}licke for enlightening
discussions and J.W. Cochrane for technical support.
R.D. and A.P.M. acknowledge additional support of ARC postdoctoral
grants. M.Y.S. and A.R.H. acknowledge additional support of
an ARC Federation grant and an ARC Professorial grant
respectively. This work was funded by the ARC and the EPSRC.


\begin{thebibliography}
\bibitem{note2} Note that the reduction in the $g$-factor anisotropy
for the 0.7 structure does not mean that the behavior of either $g^{*}_{\parallel}$ or $g^{*}_{\perp}$ in non-monotonic,
but that $g^{*}_{\parallel}$ and $g^{*}_{\perp}$ may be changing at different rates as the wire is made narrower.
\end{thebibliography}
\end{document}